\documentclass[%
 preprint,
 nofootinbib,
 nobibnotes,
 onecolumn
 prd,
]{revtex4-1}
\usepackage{tabularx}
\usepackage{graphicx}
\usepackage{subfigure}
\usepackage{dcolumn}
\usepackage{bm}
\usepackage{hyperref}
\usepackage{epstopdf}
\usepackage{amssymb}
\usepackage{float}
\usepackage[T1]{fontenc}
\usepackage{ae,aecompl,times}
\usepackage{color}
\usepackage[normalem]{ulem}

\begin{document}

\title{Can decaying vacuum elucidate the late-time dynamics of the Universe ?}

\author{Yang-Jie Yan$^{1}$}
 \email{yan_yj@mail.nankai.edu.cn}
\author{Deng Wang$^{2}$}
 \email{Cstar@mail.nankai.edu.cn}
\author{Xin-He Meng$^{1}$}
 \email{xhm@nankai.edu.cn}

\affiliation{
  $^1${Department of Physics, Nankai University, Tianjin 300071, China}\\
 $^2${Theoretical Physics Division, Chern Institute of Mathematics, Nankai University, Tianjin 300071, China}\\}

\begin{abstract}
We examine the decay vacuum model with a parameter $\epsilon$ that indicates the vacuum energy decay rate. By constraining this model with cosmic microwave background radiation, baryon acoustic oscillation, type Ia supernovae and 30 H(z) cosmic chronometer data points, we find that $\epsilon=-0.0003\pm0.00024$ with the best fitted $\chi^{2}$ value slightly smaller than that in the $\Lambda$CDM model. A negative value of $\epsilon$ suggesting dark matter energy decay into vacuum energy. We also obtain the Hubble constant $H_{0}=68.05\pm0.56$ that can alleviate the current $H_{0}$ tension between the local observation by the Hubble Space Telescope and the global measurement by the Planck Satellite. Using the effective equation of state formalism, we find this model is quintessence-like.

\end{abstract}

\maketitle


\section{\label{sec:intro}Introduction}
In the past almost two decades, a large number of observational evidences indicate that our Universe is undergoing a phase of accelerated expansion \cite{Riess1998, Perlmutter1999, Weinberg2013, PlanckCollaboration2013}. Dark energy (DE) is introduced to understand this phenomenon of late-time accelerating expansion. The most popular candidate of dark energy is the cosmological constant (CC) or $\Lambda$ term \cite{Carroll1992} in the framework of general relativity (GR). The $\Lambda$-cold-dark-matter ($\Lambda$CDM) model can explain the current cosmological observations very well, but there are two unsolved puzzles, i.e., the so-called fine-tuning \cite{weinberg1989} and coincidence problems \cite{Ostriker1995}. The former indicates that the $\Lambda$ value inferred by observations ($\rho_{\Lambda}=\Lambda/8\pi G\lesssim10^{-47} GeV^{4}$) differs from theoretical estimates given by quantum field theory ($\rho_{\Lambda}\sim10^{71} GeV^{2}$) by almost 120 orders of magnitude, while the latter, which implies that the problem with $\Lambda$ is to understand why dark energy density is not only small, but also of the same order of magnitude of the energy density of cold dark matter (CDM). As a consequence, to alleviate or even solve these problems, a flood of dark energy models are proposed and studied by cosmologists, such as phantom \cite{Caldwell2002}, quintessence \cite{Fujii1982, Ford1987, WETTERICH1988668, Hebecker2001}, decaying vacuum \cite{Wang2005, Lima1996dv, Lima1999}, bulk viscosity \cite{Meng2007, Ren2006, Ren:2005nw, Wang2017b}, Chaplygin gas \cite{Kamenshchik2001,Bento2002} and so on.

In this study, we focus on decaying vacuum model (DVM), which attempts to alleviate the coincidence problem by allowing the CDM and DE to interact with each other \cite{Ferreira2012}. We follow the model proposed by Wang and Meng that assume the form of the modification of the CDM expansion rate due to the DV effect. This decaying vacuum scenario has been discussed, for example in Ref. \cite{Alcaniz2005} the authors used SNe Ia, Chandra measurements of the X-ray gas mass fraction in 26 galaxy clusters and CMB data from WMAP and get the vacuum decay rate parameter lies on the interval $\epsilon=0.11\pm0.12$. Ref. \cite{Jesus2008} have shown that the constraint of decay rate parameter $\epsilon=0.000_{-0.000}^{+0.057}$ after using SNe Ia, BAO and CMB shift parameter data sets. In this work, we plan to re-examine this DVM by using the latest observational data and use the \textbf{CAMB} and \textbf{CosmoMC} \cite{Lewis1999bs,Lewis2002ah} packages with the Markov Chain Monte Carlo (MCMC) method.

This paper is organized as follows. In Sect. \ref{sec:model}, we review briefly the model that proposed by Wang and Meng. In Sect. \ref{sec:obsercations and numerical calculations}, we constrain this model by using the recent cosmological observations and analyze the results we obtain. The conclusions are presented in Sect. \ref{sec:summary}.

\section{decaying vacuum model}
\label{sec:model}
Here, we follow the arguments exhibited in Ref. \cite{Wang2005}, where the standard continuity equation for CDM has been modified,
\begin{eqnarray}
\label{eq.1}
  \dot{\rho}_{m}+3H\rho_{m} &=& -\dot{\rho}_{\Lambda},
\end{eqnarray}
where $\rho_{\Lambda}$ and $\rho_{m}$ are the energy density of vacuum and CDM, respectively.

Since consider the interaction between vacuum energy and CDM, there is a deviation of CDM's evolution from the standard evolution, which can characterize by a term $\epsilon$, i.e.,
\begin{eqnarray}
\label{eq.2}
  \rho_{m} &=& \rho_{m0}a^{-3+\epsilon},
\end{eqnarray}
where $\rho_{m0}$ is the current CDM energy density, $\epsilon$ is a constant parameter that describe the vacuum energy decay rate. If $\epsilon>0$ implies vacuum energy decay into CDM, thus the CDM component will dilute more slowly. On the contrary, $\epsilon<0$ implies CDM decay into vacuum. And $\epsilon=0$ corresponding to non-interacting scenario.

Now, by integrating Eq. \ref{eq.1} one can get
\begin{eqnarray}
\label{eq.3}
  \rho_{\Lambda} &=& \tilde{\rho}_{\Lambda 0}+\frac{\epsilon \rho_{m0}}{3-\epsilon}a^{-3+\epsilon},
\end{eqnarray}
where $\tilde{\rho}_{\Lambda 0}$ is an integration constant representing the ground state of the vacuum. Note that, for this DVM, we have ignored the contribution from the radiation and spatial curvature components, and considered the physical equation of stat (EOS) of the vacuum $\omega_{\Lambda}\equiv p_{\Lambda}/\rho_{\Lambda}$ equal to constant $-1$. So the Friedmann equation can be rewritten as

\begin{equation}\label{eq.4}
 \frac{H^{2}}{H_{0}^{2}}=(\frac{3 \Omega_{m0}}{3-\epsilon}a^{-3+\epsilon}+1-\frac{3\Omega_{m0}}{3-\epsilon}).
\end{equation}

\section{numerical calculations and analysis results}
\label{sec:obsercations and numerical calculations}

\begin{figure}
  \centering
  \includegraphics[scale=0.6]{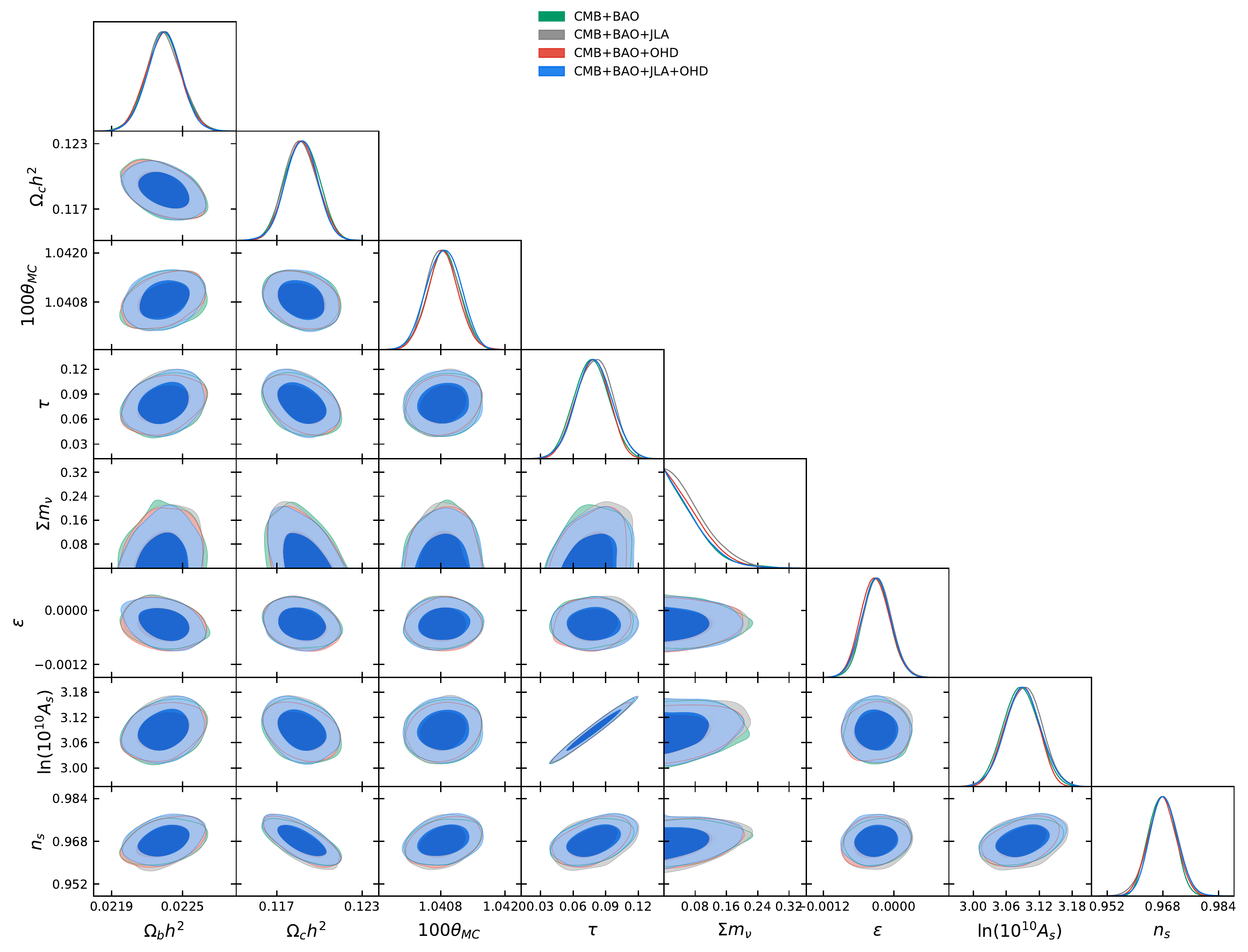}
  \caption{The one dimensional distributions on the individual parameters and two dimensional marginalized contours of the DVM, where the contour lines represent $68\%$ and $95\%$ C. L., respectively. }
  \label{fig:triangle}
\end{figure}

To study quantitatively the properties of dark energy, we perform global constraints on our DVM by using the latest cosmological observations, which are exhibited as follows: (i) \emph{CMB}: the CMB temperature and polarization data from \emph{Planck 2015} \cite{PlanckCollaboration2015}, which includes  the likelihoods of temperature (TT) at $30 \leqslant l \leqslant2500$, the cross-correlation of temperature and polarization (TE), the polarization (EE) power spectra, and the Planck low-$l$ temperature and polarization likelihood at $2\leqslant l \leqslant29$. (ii) \emph{BAO}: we employ four BAO data points: Six Degree Field Galaxy Survey (6dFGS) sample at effective redshift $z_{eff}=0.106$ \cite{Beutler2011}, the SDSS main galaxy sample (MGS) at $z_{eff}=0.15$ \cite{Ross2015}, and the LOWZ at $z_{eff}=0.32$ and CMASS $z_{eff}=0.57$ samples of the SDSS-III BOSS DR12 sample \cite{Cuesta2016}. (iii) \emph{JLA}: the "Join Light-curve Analysis" (JLA) sample of Type Ia supernova \cite{Betoule2014}, used in this paper, is constructed from the SNLS, SDSS, HST and several samples of low-$z$ SN. This sample consists of 740 SN Ia data points covering the redshift range $z \in [0.01,1.3]$. (iv) \emph{OHD}: the observational Hubble parameter data with 30 point \cite{Moresco2016a,Wang2017,Geng2017a}.

We adopt the Markov Chain Monte Carlo (MCMC) method with the data above mention to constrain this DVM. We modifies the public package \textbf{CosmoMC} and Boltzmann code \textbf{CAMB} to infer the posterior probability distributions of cosmological parameters.  In addition, the $\chi^{2}$ function for the data from $H(z)$ is take to be
\begin{equation}\label{}
  \chi^{2}=\sum_{i=1}^{n}\frac{(H_{th}(z_{i})-H_{o}(z_{i}))^{2}}{E^{i}}
\end{equation}
where the $H_{th}$ is the theoretical prediction, calculated from CAMB, and $H_{o}$ and $E$  represent the observational value and error, respectively.

\begin{table}[!hbp]
\newcommand{\tabincell}[2]{\begin{tabular}{@{}#1@{}}#2\end{tabular}}
\caption{The prior ranges, the best-fitting values and 1$\sigma$ marginalized uncertainties of cosmological parameters of the DVM, and the numbers in the bracket represent the best-fit values of the $\Lambda$CDM model. }\label{Table:parameters}
\scalebox{0.9}{
\begin{tabular}{|c|c|c|c|c|c|}
\hline
Parameters& Priors&
 \tabincell{c}{CMB+BAO} &
 \tabincell{c}{CMB+BAO+JLA}&
 \tabincell{c}{CMB+BAO+OHD}&
 \tabincell{c}{CMB+BAO\\+JLA+OHD}   \\
\hline
{\boldmath$\epsilon $} & $[-0.3, 0.3]$   &
 $-0.00029\pm0.00023$ &
 $-0.00028\pm 0.00024$&
 $-0.00032\pm 0.00024$ &
 $-0.00030\pm 0.00024$  \\
\hline
{\boldmath$\Omega_b h^2   $} & $[0.005, 0.1]$  &
 \tabincell{c}{$0.02234\pm 0.00015$\\$(0.02233)$} &
 \tabincell{c}{$0.02234\pm 0.00014$\\(0.02232)} &
 \tabincell{c}{$0.02234\pm 0.00015$\\(0.02233)} &
 \tabincell{c}{$0.02234\pm 0.00014$\\(0.02232)}      \\
\hline
{\boldmath$\Omega_c h^2   $} & $[0.001, 0.99]$ &
 \tabincell{c}{$0.1188\pm 0.0012$\\(0.1181)} &
 \tabincell{c}{$0.1187\pm 0.0011$\\(0.1182)} &
 \tabincell{c}{$0.1187\pm 0.0011$\\(0.1182)} &
 \tabincell{c}{$0.1187\pm 0.0011$\\(0.1181)}      \\
\hline
{\boldmath$\Sigma m_\nu   $} & $[0,5]$ &
 $<0.0781(<0.0787)$ &
 $<0.0859(<0.0689)$ &
 $<0.0809(<0.0763)$ &
 $<0.0773(<0.0698)$        \\
\hline
{\boldmath$n_s $} & $[0.8, 1.2]$    &
 \tabincell{c}{$0.9677\pm 0.0036$\\(0.9698)} &
 \tabincell{c}{$0.9680\pm 0.0042$\\(0.9693)} &
 \tabincell{c}{$0.9678\pm 0.0039$\\(0.9696)} &
 \tabincell{c}{$0.9683\pm 0.0040$\\(0.9698)}      \\
\hline
{\boldmath$H_{0}$} & $[20,100]$ &
 \tabincell{c}{$67.85_{-0.52}^{+0.64}$\\(68.02)} &
 \tabincell{c}{$67.82\pm 0.55$\\(68.03)} &
 \tabincell{c}{$67.84\pm 0.55$\\(68.00)} &
 \tabincell{c}{$67.87_{-0.50}^{+0.55}$\\(68.06)}      \\
\hline
{\boldmath$\Omega_{m}$}     & -  &
 \tabincell{c}{$0.3081_{-0.0082}^{+0.0068}$\\(0.3050)} &
 \tabincell{c}{$0.3083\pm 0.0069$\\(0.3049)} &
 \tabincell{c}{$0.3081\pm 0.0069$\\(0.3054)} &
 \tabincell{c}{$0.3078\pm 0.0066$\\(0.3046)}      \\
 \hline
{\boldmath$\sigma_{8}$}      & - &
 \tabincell{c}{$0.826_{-0.013}^{+0.018}$\\(0.825)} &
 \tabincell{c}{$0.826_{-0.014}^{+0.017}$\\(0.827)} &
 \tabincell{c}{$0.826_{-0.014}^{+0.017}$\\(0.827)} &
 \tabincell{c}{$0.828_{-0.014}^{+0.018}$\\(0.828)}      \\
 \hline
{\boldmath$z_{\rm{eq}}$}      & - &
 \tabincell{c}{$3372\pm 26$\\(3355)} &
 \tabincell{c}{$3370_{-24}^{+27}$\\(3357)} &
 \tabincell{c}{$3371\pm 25$\\(3358)} &
 \tabincell{c}{$3371\pm 25$\\(3356)}      \\
 \hline
{\boldmath$\chi_{min}^{2}$}  & -  &
 12980.634(12985.332) &
 13678.974(13682.862) &
 12997.854(12998.46) &
 13694.348(13696.51) \\
\hline
\end{tabular}}
\end{table}

\begin{figure}
  \centering
  \includegraphics[width=16cm]{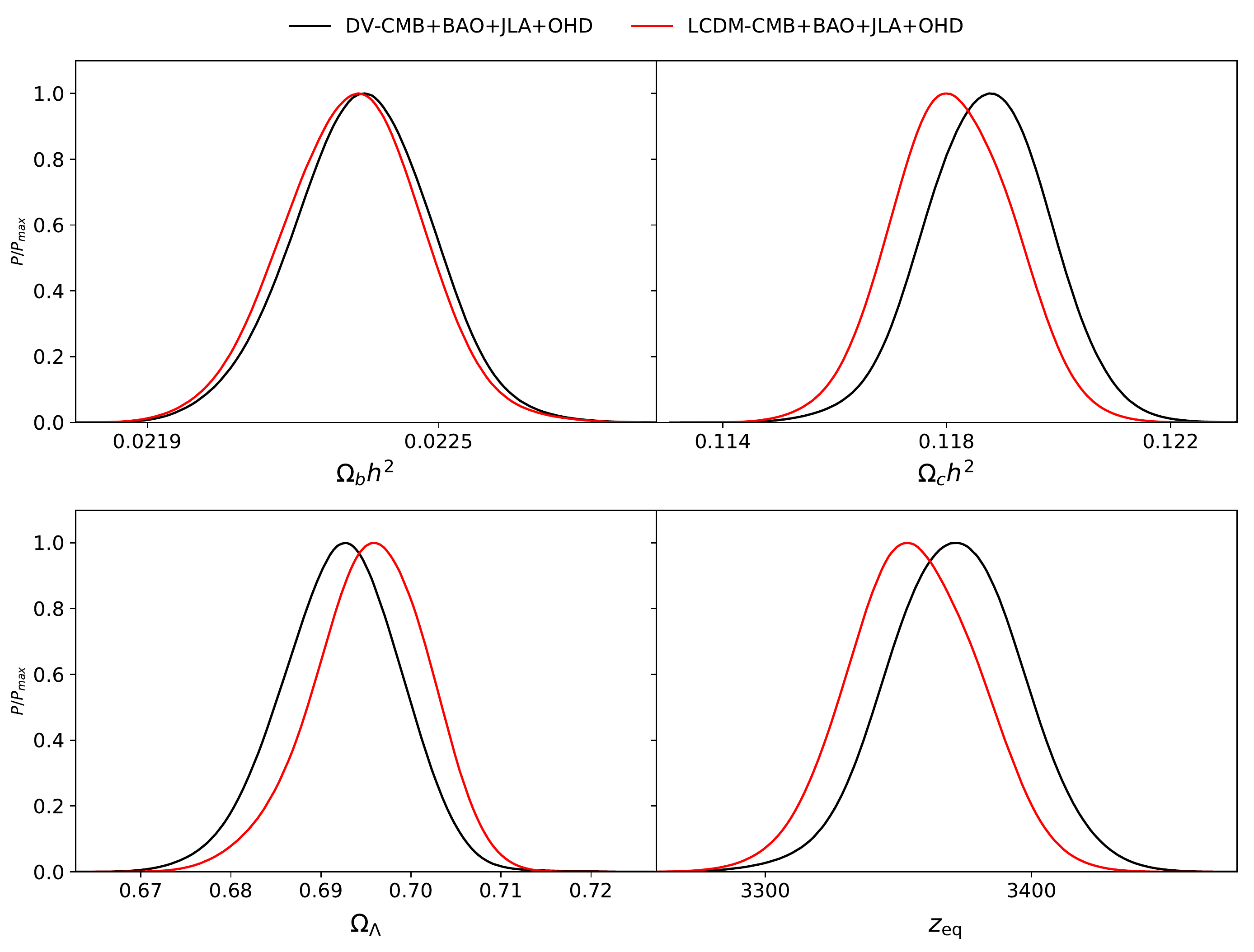}
  \caption{The 1-dimensional posterior distributions of $\Omega_{b}h^{2}$, $\Omega_{c}h^{2}$, $\Omega_{\Lambda}$ and $z_{eq}$ in the DVM(black) and $\Lambda$CDM model(red) using the data combination CMB+BAO+JLA+OHD, respectively. }
  \label{fig:1d}
\end{figure}

In Table \ref{Table:parameters} and Fig. \ref{fig:triangle}, we show the values in the best-fit points and corresponding 1$\sigma$ errors of individual parameters and 1-dimensional posterior distributions on the individual parameters and 2-dimensional marginalized contours of the DVM. One can find that the best fit value of $\epsilon$ is about $\epsilon\sim -0.0003$, this value very close to the standard non-interacting case but slightly smaller than 0, which implies CDM decay to vacuum energy, and noting that the $\chi_{VDM}^{2}\lesssim\chi_{\Lambda CDM}^{2}$ in all the datasets. We conclude that the DVM deviates slightly from the $\Lambda$CDM model and is favored by the current observations.

To study the details of constrained parameters further and compare them in the DVM and non-interacting case, i.e. $\Lambda$CDM, we exhibit their 1-dimensional posterior distributions in Fig. \ref{fig:1d}. We find that the value of baryon density $\Omega_{b}h^{2}$ of DVM almost the same as that of $\Lambda$CDM model, due to baryon matter is not involved in the interaction. Furthermore the value of CDM density $\Omega_{c}h^{2}$ and redshift of the radiation-matter equality $z_{eq}$ of the DVM are larger than those of the $\Lambda$CDM model, and the value of vacuum density $\Omega_{\Lambda}$ is smaller than those of the $\Lambda$CDM model. This result is consistent with the conclusion that CDM decaying to vacuum energy.

\begin{figure}[ht]
\begin{minipage}{0.47\linewidth}
  \centerline{\includegraphics[width=7.5cm]{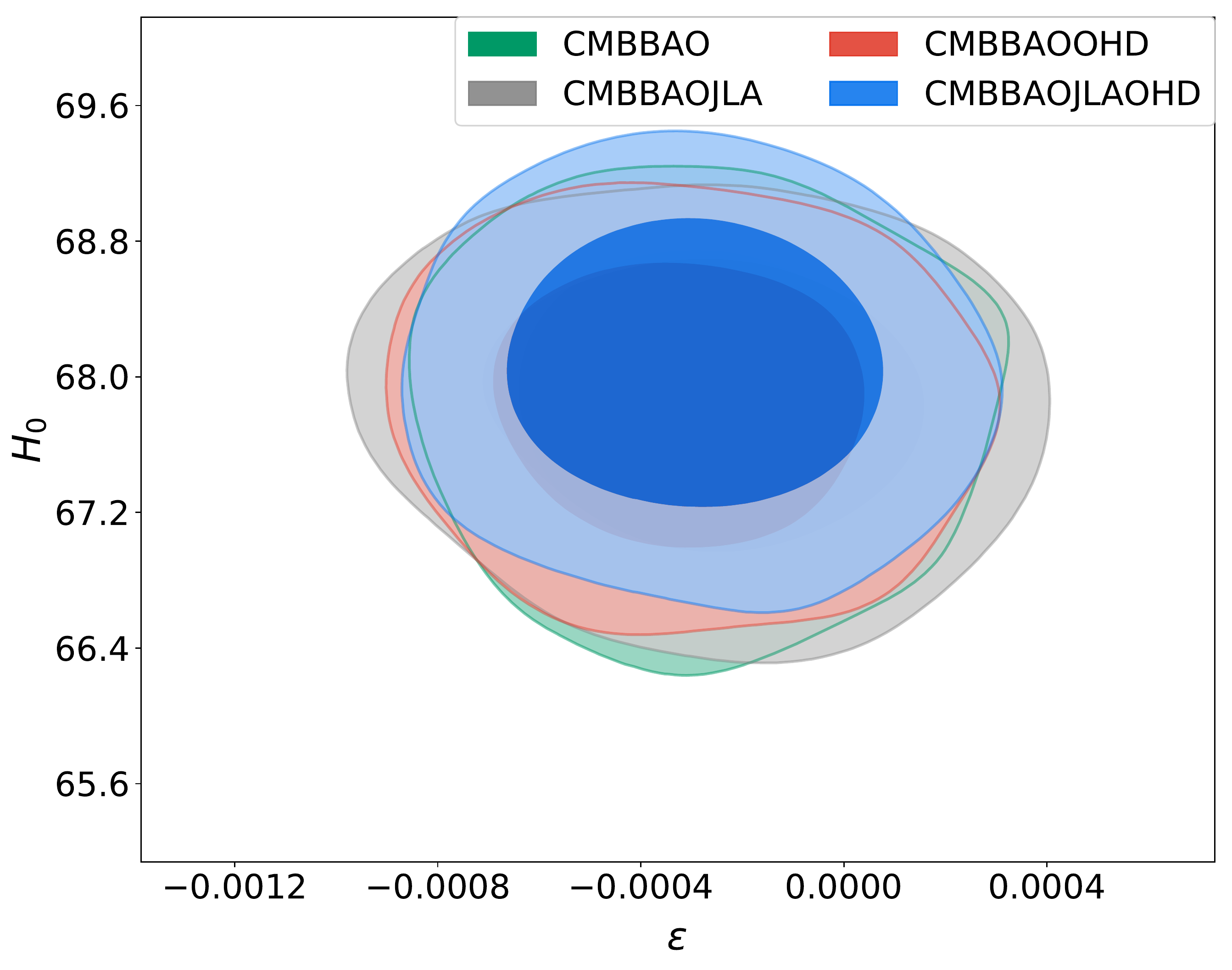}}
\end{minipage}
\hfill
\begin{minipage}{.47\linewidth}
  \centerline{\includegraphics[width=7.5cm]{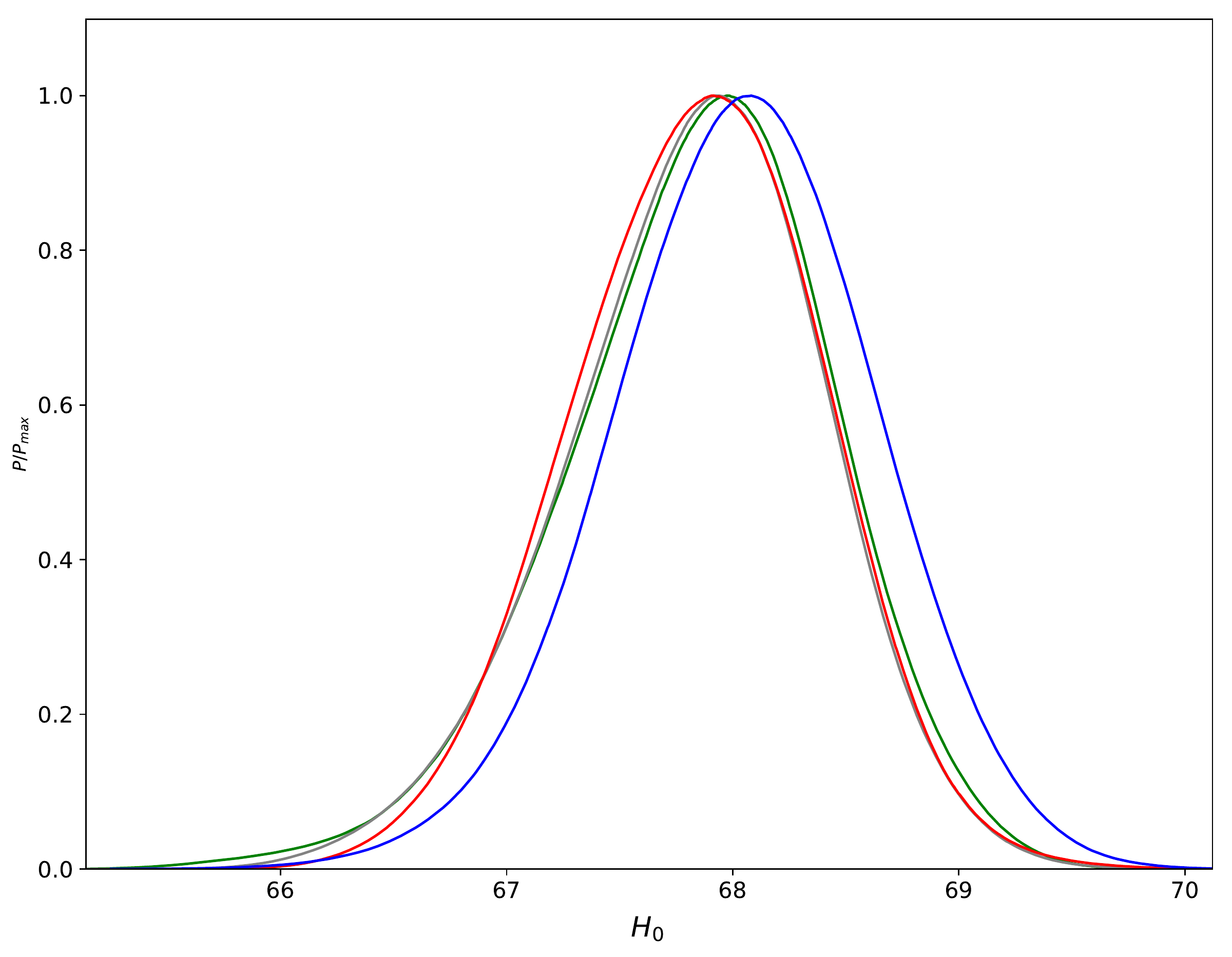}}
\end{minipage}
\caption{The two dimensional marginalized contour in the $H_{0}-\epsilon$ plane and 1-dimensional posterior distribution of $H_{0}$ }
\label{fig:H0}
\end{figure}

We also interested in the tension between the improved local measurement $H_{0}=73.24\pm1.74\textrm{km s}^{-1}\textrm{Mpc}^{-1}$ from Riess et al. \cite{Riess2016} with the Planck 2015 release $H_{0}=66.93\pm0.62\textrm{km s}^{-1}\textrm{Mpc}^{-1}$ \cite{PlanckCollaboration2015}. Using the data combination CMB+BAO+JLA+OHD, in fig. \ref{figure:H0}, we obtain $H_{0}=67.87_{-0.5}^{+0.55}\textrm{km s}^{-1}\\\textrm{Mpc}^{-1}$, and the $H_{0}$ tension can be alleviated from $3.41\sigma$ to $2.95\sigma$.

\begin{figure}
  \centering
  \includegraphics[width=10cm]{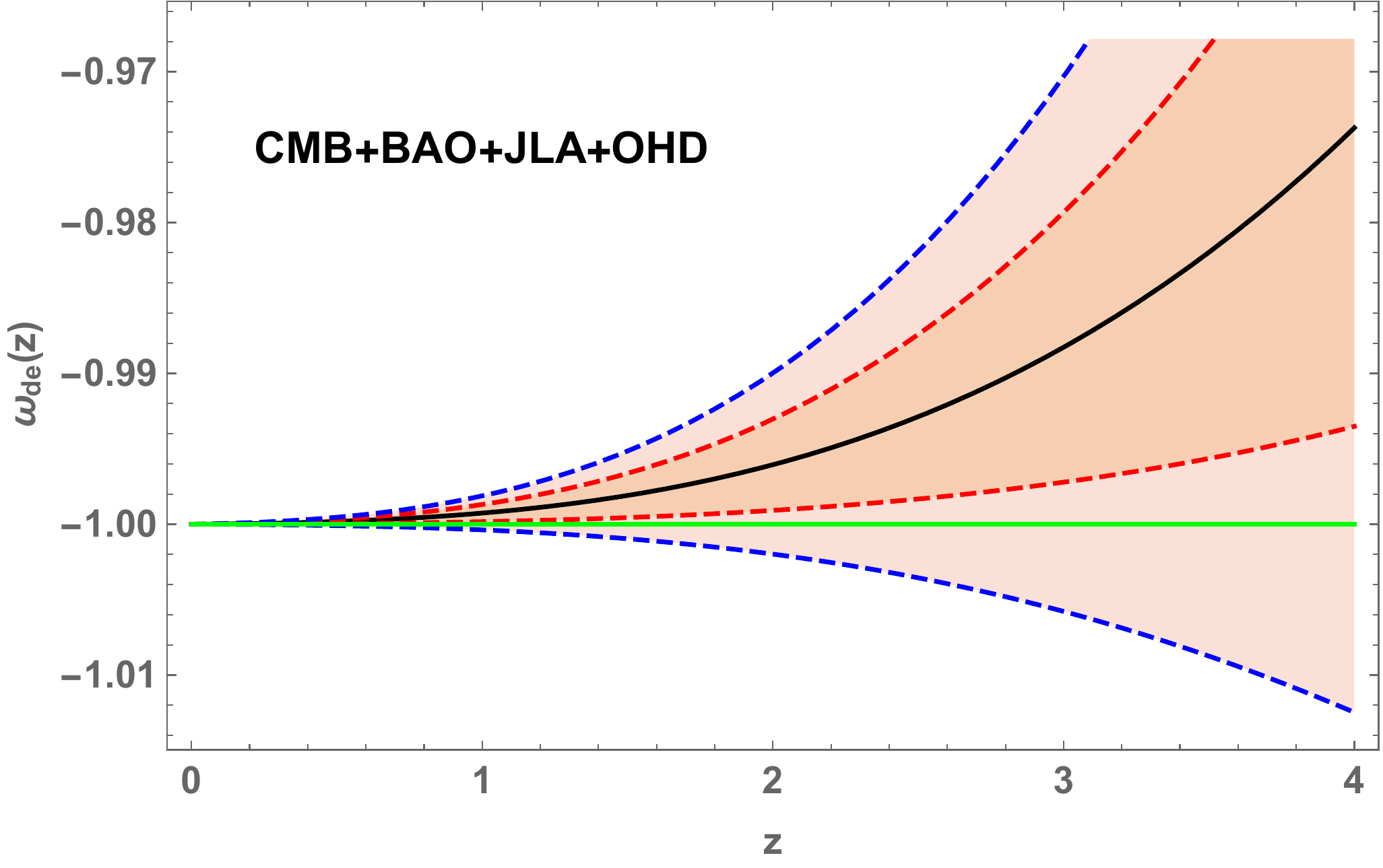}
  \caption{The relation between the redshift and the effective EoS of vacuum using combination CMB+BAO+JLA+OHD. The black and green (solid) lines correspond to the VD model and $\Lambda$CDM model, respectively. The orange and pink regions between the red and blue (dashed) lines are represent the $1\sigma$ and $2\sigma$ regions, respectively.}\label{fig:wofz}
\end{figure}

Additionally, we calculate the effective EOS of vacuum \cite{Wang2005}:
\begin{equation}\label{eq.EOS}
  \omega_{x}=-1+\frac{(1+z)^{3-\epsilon}-(1+z)^{3}}{\frac{3}{3-\epsilon}(1+z)^{3-\epsilon}-(1+z)^{3}+\frac{\tilde{\Omega_{\Lambda0}}}{\Omega_{m0}}}.
\end{equation}
In figure \ref{fig:wofz}, we show the relation between the redshift $z$ and the effective EoS of vacuum. One can find that the EoS is large then -1, denote the DVM is a quintessence-like, but get a value below -1 at 2$\sigma$ C.L., then the DVM become a phantom-like.

\section{conclusions}
\label{sec:summary}
In this work, we have reviewed the decay vacuum model. We have shown that the tightest constrain we can give, the numerical analysis results are exhibited in Table \ref {Table:parameters} and figs. \ref{fig:triangle}, \ref{fig:1d}, \ref{fig:H0}. Combining  the data of CMB temperature fluctuation and polarization, the baryon acoustic oscillations (BAO), the SN Ia sample "Joint Light-curve Analysis" (JLA) and $H(z)$ measurements, we have found that $\epsilon=-0.0003\pm0.00024$. This result is consistent with Ref. \cite{Jesus2008}, but the decay rate parameter $\epsilon$ is slightly smaller than 0 in our work, which imply the dark matter energy decay to vacuum energy. The 1-dimensional posterior distributions of $\Omega_{b}h^{2}$, $\Omega_{c}h^{2}$, $\Omega_{\Lambda}$ and $z_{eq}$ constrained by the combination CMB+BAO+JLA+OHD are supporting this conclusion. Furthermore, we find the DVM can alleviate the current $H_{0}$ tension from $3.4\sigma$ to $2.95\sigma$. Finally, we shown the effective EOS defined as equation \ref{eq.EOS}, which denote our Universe is a quintessence-like and can become a phantom case at 2 $\sigma$ C.L..

\section{acknowledgements}
YangJie Yan thanks Lu Yin for helpful communications and programming. This study is supported in part by the National Science Foundation of China.

%

\end{document}